\documentclass[1p]{elsarticle}
\usepackage{hyperref}
\usepackage{amsfonts}
\usepackage{amssymb}
\usepackage{amsmath}
\usepackage{graphicx}
\usepackage{makeidx}
\usepackage{subfig}

\journal{Chaos, Solitons, \& Fractals}
\DeclareMathOperator{\sech}{sech}
\DeclareMathOperator{\arccosh}{arccosh}

\begin{document}

\begin{frontmatter}
	
	\title{Embedded solitons in second-harmonic-generating lattices}

	\author{H.\ Susanto}
	\address{Department of Mathematics, Khalifa University, Abu Dhabi Campus, PO Box
		127788,\\ Abu Dhabi, United Arab Emirates}
	\address{Department of Mathematical Sciences, University of Essex, Wivenhoe Park,\\
		Colchester CO4 3SQ, United Kingdom}
	\ead{hadi.susanto@ku.ac.ae}

	\author{Boris A.\ Malomed}
	\address{Department of Physical Electronics, School of Electrical Engineering,
		Faculty of Engineering, and Center for Light-Matter Interaction, Tel Aviv
		University, Tel Aviv 69978, Israel}
	\address{Instituto de Alta Investigaci\'{o}n, Universidad de Tarapac\'{a}, Casilla
		7D, Arica, Chile}
	
	\begin{abstract}
		Embedded solitons are exceptional modes in nonlinear-wave systems with the propagation constant falling in the system's propagation band. An especially challenging topic is seeking for such modes in nonlinear dynamical lattices (discrete systems). We address this problem for a system of coupled discrete equations modeling the light propagation in an array of tunnel-coupled waveguides with a
		combination of intrinsic quadratic (second-harmonic-generating) and cubic
		nonlinearities. Solutions for discrete embedded solitons (DESs) are
		constructed by means of two analytical approximations, adjusted, severally, for broad and narrow DESs, and in a systematic form with the help of numerical calculations. DESs of several types, including ones with identical and opposite signs of their fundamental-frequency and second-harmonic components, are produced. In the most relevant case of narrow DESs, the
		analytical approximation produces very accurate results, in comparison with the numerical findings. In this case, the DES branch extends
from the propagation band into a semi-infinite gap as a family of regular discrete solitons. The study of stability of DESs
demonstrates that, in addition to ones featuring the well-known property of semi-stability, there are linearly stable DESs which
are genuinely robust modes.
		
	\end{abstract}
	
	
\end{frontmatter}

\section{Introduction}

One of basic scenarios of nonlinear optics is the generation of
second-harmonic (SH) waves by the fundamental-frequency (FF) input in
waveguides with a quadratic, alias $\chi ^{(2)}$, material response \cite%
{book}. The interplay of the $\chi ^{(2)}$ nonlinearity with linear paraxial
diffraction of light in planar waveguides gives rise to spatial solitons of
the $\chi ^{(2)}$ type, which are built of quadratically coupled FF and SH
components. Such solitons were first predicted as particular exact solutions
of a system of two coupled equations for the paraxial propagation of
amplitudes of the FF and SH waves \cite{Sukhorukov}, and later created
experimentally \cite{Torruellas}. Theoretical and experimental studies of $%
\chi ^{(2)}$ solitons have grown into a vast research area \cite%
{review1,review2,review3}.

Varieties of $\chi ^{(2)}$ dynamical effects and, in particular, soliton
modes, are additionally expanded in settings which combine quadratic and
cubic ($\chi ^{(3)}$) terms \cite{chichi1,chichi2}. There are two distinct
types of such combinations: competing, if the $\chi ^{(3)}$ term is
self-defocusing, and cooperating, if the latter term has the focusing sign.
One of specific phenomena predicted in models with the combined $\chi
^{(2)}:\chi ^{(3)}$ nonlinearity is the existence of \textit{embedded
solitons} \cite{yang99} i.e., localized modes whose propagation constant,
rather than belonging, as usual, to spectral gaps, in which the propagation
of linear waves is suppressed by the system's dispersion relation, falls (is
\textit{embedded})\ in a propagation band. Normally, solitons cannot exist
in such a case, as the resonance between propagation constants of the
solitons and linear waves gives rise to decay of the (quasi-) soliton modes
into radiation. Nevertheless, in exceptional cases. i.e., at some isolated
values of the soliton's propagation constant, the decay rate may vanish (in
other words, the resonance strength becomes exactly equal to zero), which
explains the existence of embedded solitons. This possibility was originally
proposed in Refs. \cite{Champneys} and \cite{Fujioka} (without naming such
modes \textquotedblleft embedded" ones), see an early review of the topic in
Ref. \cite{ESreview}. In the linear version of the $\chi ^{(2)}$ system,
dispersion relations for the FF and SH waves are decoupled, each consisting
of a semi-infinite propagation band and forbidden gap. Boundaries between
the gap and band in the FF and SH spectra are mutually shifted, in the
general case, due to the mismatch between the FF and SH components. In terms
of the FF wave, the propagation constant of any soliton must always belong
to the respective gap. However, the SH propagation equation, with the SH
amplitude driven by the square of the FF, may have \textit{nonlinearizable}
solutions. For such states, the quadratic (FF$\times $FF) term in the SH
equation cannot be omitted when considering exponentially decaying tails of
the soliton's fields, as it has the same order of magnitude as linear SH
terms (\textquotedblleft tail-locked" modes) \cite{yang99}. Thus, the SH
propagation constant of these solutions does not need to obey the spectral
structure of the linearized equation, falling, counter-intuitively, into the
propagation band for the SH component.

Because, as mentioned above, embedded solitons exists only at specially
selected value(s) of the propagation constant, they do not form continuous
families, being categorized, accordingly, as \textit{codimension-one}
solutions (see, e.g. Ref. \cite{codim}). Another specific feature of
embedded solitons is their (in)stability. In most cases, they are neutrally
stable against small perturbations in the linear approximation, but are
unstable against the (slower-than-exponential) growth of perturbations
driven by terms which are quadratic, with respect to the small perturbations
\cite{yang99}. In some cases, such \textquotedblleft semi-stable" solitons
may seem as practically stable ones, if the nonlinear growth of the
perturbations is extremely slow \cite{ESreview}, and there are particular
models which support completely stable embedded solitons \cite{stable}.

One of ramifications of studies of $\chi ^{(2)}$ solitons is the creation of
their discrete counterparts in arrays (lattices) of tunnel-coupled
waveguides with the quadratic or combined quadratic-cubic intrinsic
nonlinearities in individual guiding cores \cite{Stegeman,PhysRep}. A
challenging possibility is to search for discrete embedded solitons (DESs)
in the respective system of lattice-propagation equations for amplitudes of
the FF and SH waves, coupled by the quadratic nonlinearity acting at lattice
sites. This objective was a subject of Ref. \cite{Kazu}. In that work, only
a partial consideration of the problem was performed; in particular, only
DESs with opposite signs of the FF and SH components were found. Note that
no analytical methods were applied in Ref. \cite{Kazu}. In the present work,
we aim to develop the consideration of DESs in $\chi ^{(2)}$ lattices,
elaborating an analytical approximation (the averaging method), in parallel
to producing systematic numerical results, and producing DESs of more
general types, featuring both identical and opposite signs of the FF and SH
components. Additionally, we also consider stability of the DES.

The model is formulated in Section \ref{sec2}, and two analytical methods
are presented in Section \ref{sec3}. One of them (the quasi-continuum
approximation) is relevant for broad DESs, and the opposite anti-continuum
approximation, applies, with high accuracy, to strongly discrete solitons.
Numerical results are reported, and compared to the analytical
approximations, in Section \ref{sec4}. 
Stability of DES is investigated in the same section by means of computing
eigenvalues of the linearisation operator and direct simulations of the perturbed evolution. We also report the existence of linearly stable embedded solitons in the weakly coupled lattice. The paper is completed by Section \ref{conc}.

\section{The model}

\label{sec2}

We start with the discrete one-dimensional system with the on-site $\chi
^{(2)}:\chi ^{(3)}$ nonlinearity, which was introduced in Ref.\ \cite{Kazu}:
\begin{subequations}
\begin{align}
& i\frac{du_{n}}{dz}+\epsilon \Delta u_{n}+u_{n}^{\ast }v_{n}+\gamma
_{1}(|u_{n}|^{2}+2|v_{n}|^{2})u_{n}=0,  \label{uv1} \\
& i\frac{dv_{n}}{dz}-\epsilon \delta \Delta v_{n}+qv_{n}+\frac{1}{2}%
u_{n}^{2}+2\gamma _{2}(|v_{n}|^{2}+2|u_{n}|^{2})v_{n}=0.  \label{uv2}
\end{align}%
\label{gov}
\end{subequations}
Here $\epsilon $ is the coefficient of coupling between adjacent sites of
the underlying lattice for complex amplitude $u_{n}$ of the FF wave, $\delta
\gtrless 0$ is the relative lattice-coupling coefficient for amplitude $%
v_{n} $ of the SH wave, the finite-difference Laplacian is $\Delta
X_{n}=X_{n+1}+X_{n-1}-2X_{n}$, the SH-FF mismatch is accounted for by
constant $q$, the coefficient of the\ $\chi ^{(2)}$ interaction is scaled to
be $1$, and $\gamma _{1,2}$ are coefficients of the $\chi ^{(3)}$
self-interaction for the FF and SH waves. All coefficients in Eqs.\ (\ref%
{uv1}) and (\ref{uv2}) are real, while $\ast $ stands for the complex
conjugate.\

Note that, rescaling the lattice fields $u_{n}$ and $v_{n}$ by a common
factor, one can fix the value of $\gamma _{1}$, while $\gamma _{2}$ remains
an independent parameter. In this work, we use this option by fixing $%
\left\vert \gamma _{1}\right\vert =0.05$. As for $\gamma _{2}$, we produce
numerical results for $\gamma _{2}=\gamma _{1}$, as variation of ratio $%
\gamma _{2}/\gamma _{1}$ does not lead to a conspicuous change of the
results. The sign of $\gamma _{1,2}$ is essential, positive and negative
ones representing, respectively, the self-focusing and defocusing $\chi
^{(3)}$ nonlinearity. In the absence of the $\chi ^{(2)}$ terms, this sign
may be fixed by means of the \textit{staggering transformation}, $\left(
u_{n},v_{n}\right) \rightarrow \left( -1\right) ^{n}\left(
u_{n},v_{n}\right) $ \cite{staggered}. However, the present system does not
admit it, as the sign of the $\chi ^{(2)}$ coefficient is already fixed.
Below, we consider both signs of $\gamma _{1,2}$.

The FF\ lattice coupling constant in Eq.\ (\ref{uv1}) may be set to be
positive, $\epsilon >0$ (if it is originally negative, its sign may be
reversed by means of the complex conjugation of the equations). On the other
hand, $\delta >0$ or $\delta <0$ in Eq.\ (\ref{uv2}) implies that the sign
of the coupling constant in the SH subsystem is, respectively, opposite or
identical to one for the FF wave, both cases being considered below. In the
experiment, the magnitude and sign of the coupling constant in arrays of
optical waveguides may be changed by means of the known
diffraction-managment scheme \cite{manage1,manage2}.

Stationary solutions to Eqs.\ (\ref{gov}) are looked for as
\begin{equation}
u_{n}=e^{i\omega z}U_{n},~v_{n}=e^{2i\omega z}V_{n},  \label{stat}
\end{equation}%
where real $\omega $ is a propagation constant, and real discrete fields $%
U_{n}$, $V_{n}$ satisfy the following equations:
\begin{subequations}
\begin{align}
& (-\omega +\epsilon \Delta )U_{n}+U_{n}V_{n}+\gamma _{1}\left(
U_{n}^{2}+2V_{n}^{2}\right) U_{n}=0,  \label{U} \\
& \left( q-2\omega -\epsilon \delta \Delta \right) V_{n}+\frac{1}{2}%
U_{n}^{2}+2\gamma _{2}\left( V_{n}^{2}+2U_{n}^{2}\right) V_{n}=0.  \label{V}
\end{align}%
Llinearizing these equations in a formally complex form, it is
straightforward to derive dispersion relations for the plane waves $\sim
\exp \left( ikn\right) $ in the FF and SH components:
\end{subequations}
\begin{equation}
\omega =-2\epsilon (1-\cos k),\quad \omega =\epsilon \delta (1-\cos k)+q/2,
\label{disp}
\end{equation}%
where $k$ is the spatial wavenumber. As it follows from Eq.\ (\ref{disp}),
the propagation band for the plane waves in the FF component is
\begin{equation}
-4\epsilon <\omega <0,  \label{FFband}
\end{equation}%
while for the SH component it is
\begin{gather}
q/2<\omega <q/2+2\epsilon \delta ~\mathrm{for~}\delta >0,
\label{SHband-posdelta} \\
q/2-2|\delta |\epsilon <\omega <q/2~\mathrm{for}~\delta <0.
\label{SHband-negdelta}
\end{gather}

Spectral intervals not covered by the propagation bands are considered as
gaps. Thus, each band given by Eqs.\ (\ref{FFband}) and (\ref%
{SHband-posdelta}), (\ref{SHband-negdelta}) is located between two
semi-infinite gaps for the FF and SH components, respectively. The gaps
extending to $\omega \rightarrow +\infty $ and $-\infty $ may host,
severally, discrete solitons of the unstaggered and staggered types, the
latter one assuming the structure of the lattice fields $\sim (-1)^{n}$ \cite%
{staggered}. To look for DES solutions, we assume that the FF propagation
constant belongs to the semi-infinite gap corresponding to unstaggered
discrete solitons, i.e., $\omega >0$, as it follows from Eq.\ (\ref{FFband}%
). Simultaneously, it is assumed that the SH propagation constant, $2\omega $%
, falls into the corresponding propagation band given by Eqs.\ (\ref%
{SHband-posdelta}) or (\ref{SHband-negdelta}) (otherwise, one would be
dealing with regular discrete solitons, rather than DESs).

\section{Analytical approximations}

\label{sec3}

\subsection{The continuum limit ($\protect\epsilon \gg 1$)}

For $\epsilon \gg 1$, equations\ (\ref{U}) and (\ref{V}) may be approximated
by their continuous counterparts:
\begin{subequations}
\begin{align}
-\omega U+U_{xx}+& UV+\gamma _{1}\left( U^{2}+2V^{2}\right) U=-\frac{1}{%
12\epsilon }U_{xxxx}+\mathcal{O}(1/\epsilon ^{2}),  \label{Uc} \\
-2\omega V-\delta V_{xx}+& qV+\frac{1}{2}U^{2}+2\gamma _{2}\left(
V^{2}+2U^{2}\right) V_{n}=-\frac{1}{12\epsilon }V_{xxxx}+\mathcal{O}%
(1/\epsilon ^{2}).  \label{Vc}
\end{align}%
In this limit (if the fourth derivatives may be omitted), and under the
assumption that the SH component is much weaker than the FF one, i.e., $%
|V|^{2}\ll |U|^{2}$ Eqs.\ (\ref{U}), (\ref{V}) give rise to a known
single-humped embedded-soliton solution \cite{yang99}.

The variational approximation, which may be efficiently used to construct
soliton-like solutions of discrete equations \cite{Weinstein,Kaup,Robert},
does not apply to Eqs.\ (\ref{U}), (\ref{V}), as well as to their
continuous-limit counterparts (\ref{Uc}), (\ref{Vc}), because these systems
cannot be derived from Lagrangians, except for the case of $\gamma
_{1}=2\gamma _{2}$ (numerical results are produced below for the case of $%
\gamma _{1}=\gamma _{2}$, when the Lagrangian representation is not
available). Nevertheless, it is possible to apply the averaging method \cite%
{dawe13}, which uses an adopted ansatz, such as the one introduced below in
Eq.\ (\ref{solc}), as an \textquotedblleft optimal" approximate localized
solution of Eqs.\ (\ref{U}), (\ref{V}). The averaging method amounts to the
variational one if the underlying equation is Lagrangian \cite{dawe13}.

Motivated by the example of the exact embedded soliton found in Ref. \cite%
{yang99}, we assume that localized solutions of Eqs. (\ref{Uc}) and (\ref{Vc}%
) may be approximated by
\end{subequations}
\begin{equation}
U=A\sech\left( \sqrt{\omega }x\right) ,\,V=B\sech^{2}\left( \sqrt{\omega }%
x\right) .  \label{solc}
\end{equation}%
Following the averaging method, we substitute the ansatz in Eqs.\ (\ref{Uc}%
), (\ref{Vc}), multiply the first equation by $\partial U/\partial A$ and
the second one by $\partial V/\partial B$, and integrate the resulting
expressions over $-\infty <x<+\infty $, to derive the following relations
between parameters of the ansatz:
\begin{subequations}
\begin{align}
& A^{2}=\frac{1}{\gamma _{1}}\left( {2\omega -\frac{8}{5}B^{2}\gamma _{1}-B-%
\frac{7\omega ^{2}}{120\epsilon }}\right) ,  \label{vaa} \\
& A^{2}=\frac{2B\left[ 42\left( 2\delta -5\right) \omega -144B^{2}\gamma
_{2}-105q-20\epsilon ^{-1}\omega ^{2}\right] }{21(32B\gamma _{2}+5)}.
\label{vab}
\end{align}%
Approximate solutions given by Eqs.\ (\ref{vaa}), (\ref{vab}) are candidates
for embedded solitons in the case of $\delta >0$. However, they do not
select true DESs, as the actual SH component would likely include tails
which do not vanish at $x\rightarrow \pm \infty $. We therefore need to
impose an additional condition, which is the orthogonality of the localized
solution (\ref{solc}) to the tail \cite{Dave,syaf12}.

Because of the possible presence of the nonvanishing tail in the SH
component, we seek for solutions in the form of
\end{subequations}
\begin{equation}
V=V_{0}+CV_{1},  \label{VVV}
\end{equation}%
where $V_{0}$ is approximated by Eq.\ (\ref{solc}), $C$ is a constant, and
the asymptotic form of the tail has the form of $V_{1}\sim e^{\pm ikx}$ at $%
x\rightarrow \pm \infty $, where wavenumber $k$ is given by dispersion
relation following from the linearization of Eq.\ (\ref{Vc}), i.e.,
\begin{equation}
k^{2}=6\epsilon \left( -\delta \pm \sqrt{\delta ^{2}+(2\omega -q)/3\epsilon }%
\right) .  \label{k}
\end{equation}%
Making use of the averaging method once again, we multiply Eq.\ (\ref{Vc})
by $\partial V/\partial C$, substitute $C=0$ in the resulting expression (as
we want to identify a DES), and integrate the equation over $-\infty <x<%
\mathbb{+\infty }$, to derive the orthogonality condition
\begin{equation}
J\equiv \int_{-\infty }^{+\infty }N(U_{0},V_{0})\cos (kx)dx=0,  \label{solv}
\end{equation}%
where $U_{0}$ is ansatz (\ref{solc}) for the approximate localized solution,
and
\begin{equation}
N(U,V)=-2\omega V-\delta V_{xx}+qV+(1/2)U^{2}+2\gamma _{2}\left(
V^{2}+2U^{2}\right) V_{n}+\left( 12\epsilon \right) ^{-1}V_{xxxx}.
\end{equation}%
The integral in Eq.\ (\ref{solv}) can be evaluated by means of residues of
the analytically continued integrand,
\begin{equation}
\int_{-\infty }^{+\infty }N(V_{0})\cos (kx)\,dx=-2\pi \sum \text{Im}\,\text{%
Res}\left[ N(V_{0})e^{ikx}\right] ,
\end{equation}%
where the summation is to be performed over all poles that lie in the upper
complex half-plane of $x$. The analytical function written in Eq.\ (\ref%
{solc}) has infinitely many poles at imaginary values of the integration
variable, $x=\left( 1+2n\right) i\pi /\sqrt{\omega }$, $n=0,1,2,\dots $.
However, the asymptotic limit is determined solely by the contribution from
the lowest pole with $n=0$, which yields
\begin{gather}
J\approx -\frac{2k\pi }{3\omega ^{3}}\exp \left( \frac{k\pi }{2\sqrt{\omega }%
}\right) \times \left[ \frac{\gamma _{2}}{20}(k^{2}+4\omega )(k^{2}+16\omega
)B^{3}+\frac{3}{2}A^{2}\omega ^{2}\right.  \notag \\
\left. +2\omega ^{2}B\left( -3\omega +\left( 4A^{2}\gamma _{2}+\frac{3}{2}%
\delta k^{2}+\frac{3}{2}q\right) \right. \right. \left. \left. +\frac{1}{%
\omega }A^{2}k^{2}\gamma _{2}+\frac{1}{8\epsilon }k^{4}\right) \right] .
\label{Jc}
\end{gather}

Thus, Eqs.\ (\ref{vaa}), (\ref{vab}), (\ref{k}) and (\ref{Jc}) predict
values of parameters for DESs in the continuum limit, $\epsilon \gg 1$.

\subsection{The anti-continuum limit\ ($\protect\epsilon \ll 1$)}

In the anti-continuum limit, weakly coupled solutions of Eqs.\ (\ref{U}), (%
\ref{V}) may be approximated by the simplest ansatz \cite{Weinstein},%
\begin{equation}
U_{n}=A\exp \left( -\alpha |n|\right) ,~V_{n}=B\exp \left( -2\alpha
|n|\right) ,  \label{ansatz}
\end{equation}%
where amplitudes $A$ and $B$ are free parameters, and $\alpha $ is
determined by the condition that the ansatz must agree with a solution of
the linearized version of Eq.\ (\ref{U}) for the tail of the FF component
\cite{Dave}, i.e., with the first equation of system (\ref{disp}), in which $%
\alpha =ik$ is substituted:
\begin{equation}
\alpha =\arccosh\left( 1+\omega /2\epsilon \right) .  \label{alpha}
\end{equation}%
%
%
%
%
%
%
%
Next, we determine values of $A$ and $B$ in ansatz (\ref{ansatz}), again
using the averaging method: the ansatz is substituted in Eqs.\ (\ref{U}), (%
\ref{V}), multiplying the first equation by $\partial U_{n}/\partial A$, the
second one by $\partial V_{n}/\partial B$, and performing the summation over
$-\infty <n<+\infty $, to obtain
\begin{subequations}
\begin{align}
& \left( A^{2}\gamma _{1}+B\right) F_{4}+B^{2}\gamma _{1}F_{6}+\epsilon
e^{-\alpha }-\omega /2-\epsilon =0,  \label{va1} \\
& \left( q-2\omega +\frac{A^{2}}{2B}\right) F_{4}+2B^{2}\gamma
_{2}F_{8}+4A^{2}\gamma _{2}F_{6}+\frac{2\epsilon \delta }{F_{2}}=0,
\label{va2}
\end{align}%
where $F_{n}\equiv \coth \left( n\alpha /2\right) $.

Alternatively, the value of $B$ may also be approximated by demanding that,
if the ansatz is inserted in Eq.\ (\ref{V}),
the tail of the SH component must be \textquotedblleft locked" to the
squared FF tail at $|n|\rightarrow \infty $ (i.e., the tails of both
components must jointly solve the asymptotic form of the equation):
\end{subequations}
\begin{equation}
B\approx {A^{2}}\left[ \omega -q/2+\epsilon \delta (\cosh (2\alpha )-1)%
\right] ^{-1}.
\end{equation}%
However, the latter approximation overestimates the role of the DES' tails
in comparison to its core region, which limits the applicability of the
approximation, therefore we do not employ it here.

Similar to the situation considered above, Eqs.\ (\ref{va1}) and (\ref{va2})
do not select true DESs, while the appropriate selection is provided by the
condition of the orthogonality of the soliton ansatz to the nonvanishing
tail. To this end, the ansatz including the tail is written as $V_{n}=B\exp
\left( -2\alpha |n|\right) +C\widetilde{V}_{n}$, where $C$ is a constant, $%
\widetilde{V}_{n}\sim \cos (nk)$ as $n\rightarrow \pm \infty $, and $k$ is
determined by the second equation of the system of dispersion relations \ref%
{disp}, cf. Eq.\ (\ref{VVV}). Also similar to what was done above, the
orthogonality condition is obtained by multiplying expression (\ref{V}) by $%
\widetilde{V}_{n}$ and performing the summation over $n$:%
\begin{equation}
J=\sum_{n=-\infty }^{+\infty }\cos \left( kn\right) \times \left[ \left(
q-2\omega -\epsilon \delta \Delta \right) V_{n}+\frac{1}{2}U_{n}^{2}+2\gamma
_{2}\left( V_{n}^{2}+2U_{n}^{2}\right) V_{n}\right] =0.  \label{orth}
\end{equation}%
The infinite sum in Eq.\ (\ref{orth}) can be calculated by means of the
following formula:%
\begin{equation}
\sum_{n=-\infty }^{+\infty }\cos \left( kn\right) e^{-j\alpha |n|}\equiv \mathrm{Re}%
\left\{ \sum_{-\infty }^{+\infty }e^{-j\alpha |n|+ik|n|}\right\} =G_{j},
\label{sum}
\end{equation}%
where $G_{j}=\sinh \left( j\alpha\right) /\left[ \cosh \left(j\alpha
\right) -\cos (k)e^{-j\alpha}\right] $. Thus, Eq.\ (\ref{orth}) gives rise
to the following selection condition for DES:%
\begin{equation}
J=\left[ 2B\delta \epsilon (1-\cos k)-2\omega B+qB+A^{2}/2\right]
G_{2}+2B^{3}\gamma _{2}G_{6}+4A^{2}B\gamma _{2}G_{4}=0.  \label{select}
\end{equation}%
The form of Eq.\ (\ref{select}) clearly demonstrates that that this
condition may be satisfied \emph{only} in the case of the combined $\chi
^{(2)}:\chi ^{(3)}$ nonlinearity: if the cubic terms are absent, i.e., $%
\gamma _{2}=0$, Eq.\ (\ref{select}) cannot hold.

Thus, Eqs.\ (\ref{va1}), (\ref{va2}) and (\ref{select}) predict parameter
values for DESs in the anti-continuum limit, $|\epsilon |\ll 1$.

\section{Numerical results}

\label{sec4}


The numerical solution of stationary equations\ (\ref{U}) and (\ref{V}) was
obtained by employing an iterative optimization algorithm for the solution
of nonlinear least-squares problems through the function \texttt{fsolve} of
\textsc{Matlab}. Once DES components $U_{n}$ and $V_{n}$ had been found, we
studies stability of the solution by substituting a perturbed solution, $%
u_{n}=(U_{n}+(\hat{r}_{n}+i\tilde{r}_{n})e^{\lambda z}+(\hat{r}_{n}^{\ast }+i%
\tilde{r}_{n}^{\ast })e^{\lambda ^{\ast }z})e^{i\omega z}$ and $%
v_{n}=(V_{n}+(\hat{s}_{n}+i\tilde{s}_{n})e^{\lambda z}+(\hat{s}_{n}^{\ast }+%
\tilde{s}_{n}^{\ast })e^{i\lambda ^{\ast }z})e^{2i\omega z}$ into Eqs. (\ref%
{uv1}), (\ref{uv2}) and linearizing them with respect to small perturbations
$\hat{r}_{n}$, $\tilde{r}_{n}$, $\hat{s}_{n}$ and $\tilde{s}_{n}$, arriving
at the following problem for stability eigenvalue $\lambda $:
\begin{subequations}
\begin{equation}
-\lambda \mathrm{D}\left(
\begin{array}{cc}
\hat{r}_{n} &  \\
\tilde{r}_{n} &  \\
\hat{s}_{n} &  \\
\tilde{s}_{n} &
\end{array}%
\right) =\left(
\begin{array}{cccc}
0 & L_{1,-} & 0 & U_{n} \\
L_{1,+} & 0 & U_{n}+4\gamma _{1}U_{n}V_{n} & 0 \\
0 & U_{n} & 0 & L_{2,-} \\
U_{n}+8\gamma _{2}U_{n}V_{n} & 0 & L_{2,+} &
\end{array}%
\right) \left(
\begin{array}{cc}
\hat{r}_{n} &  \\
\tilde{r}_{n} &  \\
\hat{s}_{n} &  \\
\tilde{s}_{n} &
\end{array}%
\right) ,
\label{evp}
\end{equation}%
where we have defined a diagonal matrix,
\end{subequations}
\begin{equation}
\mathrm{D}\equiv \left(
\begin{array}{cccc}
1 & 0 & 0 & 0 \\
0 & -1 & 0 & 0 \\
0 & 0 & 1 & 0 \\
0 & 0 & 0 & -1%
\end{array}%
\right) ,
\end{equation}%
and linearisation operators
\begin{align}
& L_{1,\mp }=\epsilon \Delta -\omega \mp V_{n}+\gamma _{1}\left( (2\mp
1)U_{n}^{2}+2V_{n}^{2}\right), \\
& L_{2,\mp }=-\epsilon \delta \Delta -\omega +q+2\gamma _{2}\left( (2\mp
1)V_{n}^{2}+2U_{n}^{2}\right).
\end{align}%
DES is unstable when there are eigenvalues $\lambda $ with nonzero real
parts, and linearly stable otherwise. A semi-analytical approximation for
identifying critical eigenvalues by means of the variational formulation is
possible (see, e.g., Ref. \cite{rahmi} and references therein), but we do
not pursue this direction here.

After identifying the (in)stability of DES in terms of the eigenvalues,
their perturbed evolution was investigated, simulating Eqs. (\ref{uv1}) and (%
\ref{uv2}) by means of the fourth-order Runge-Kutta method in the temporal
direction.

\subsection{Broad discrete embedded solitons}

\begin{figure}[tbph]
\centering
\subfloat[][]{\includegraphics[width=7cm]{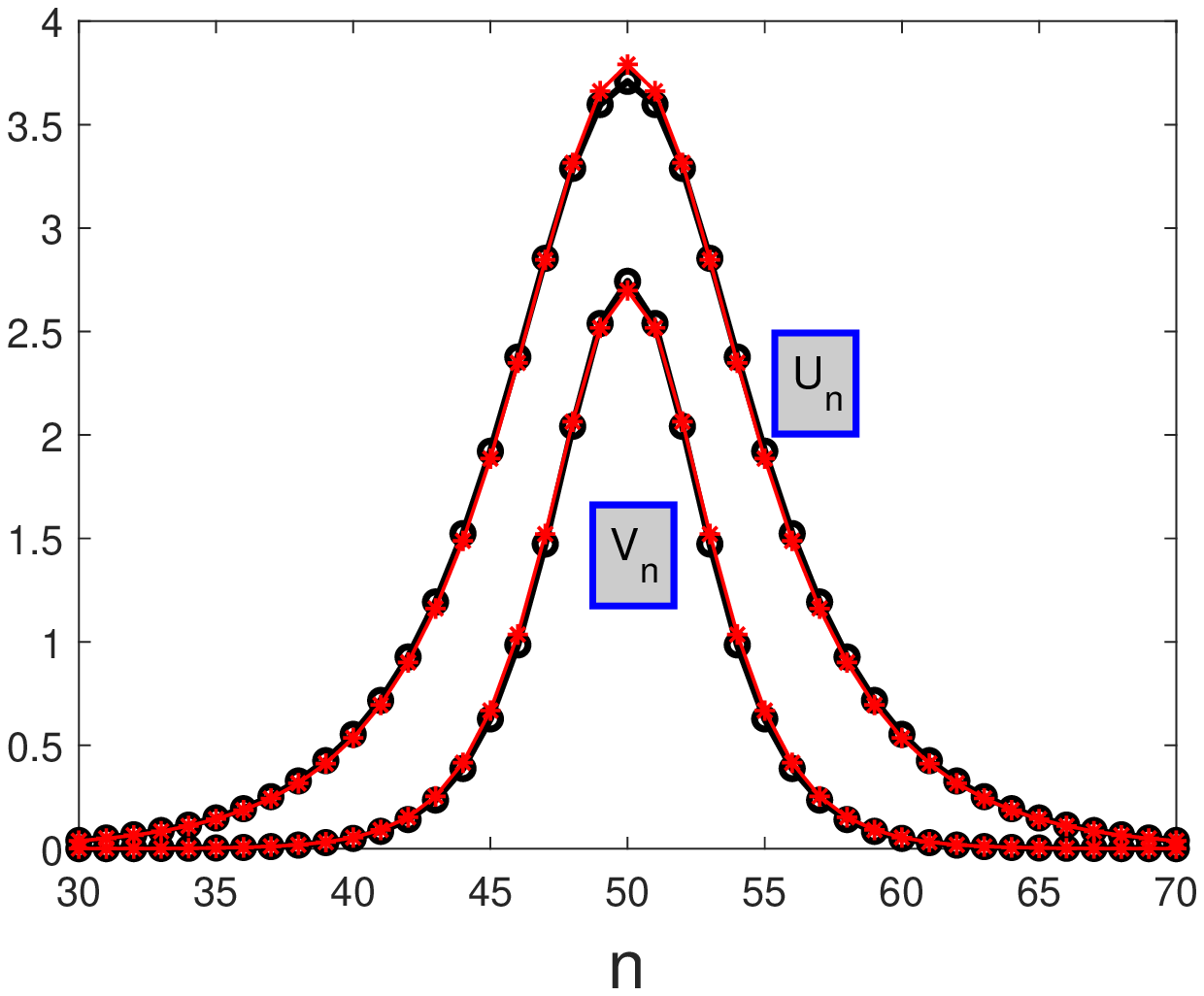}\label{fig1a}} %
\subfloat[][]{\includegraphics[width=7cm]{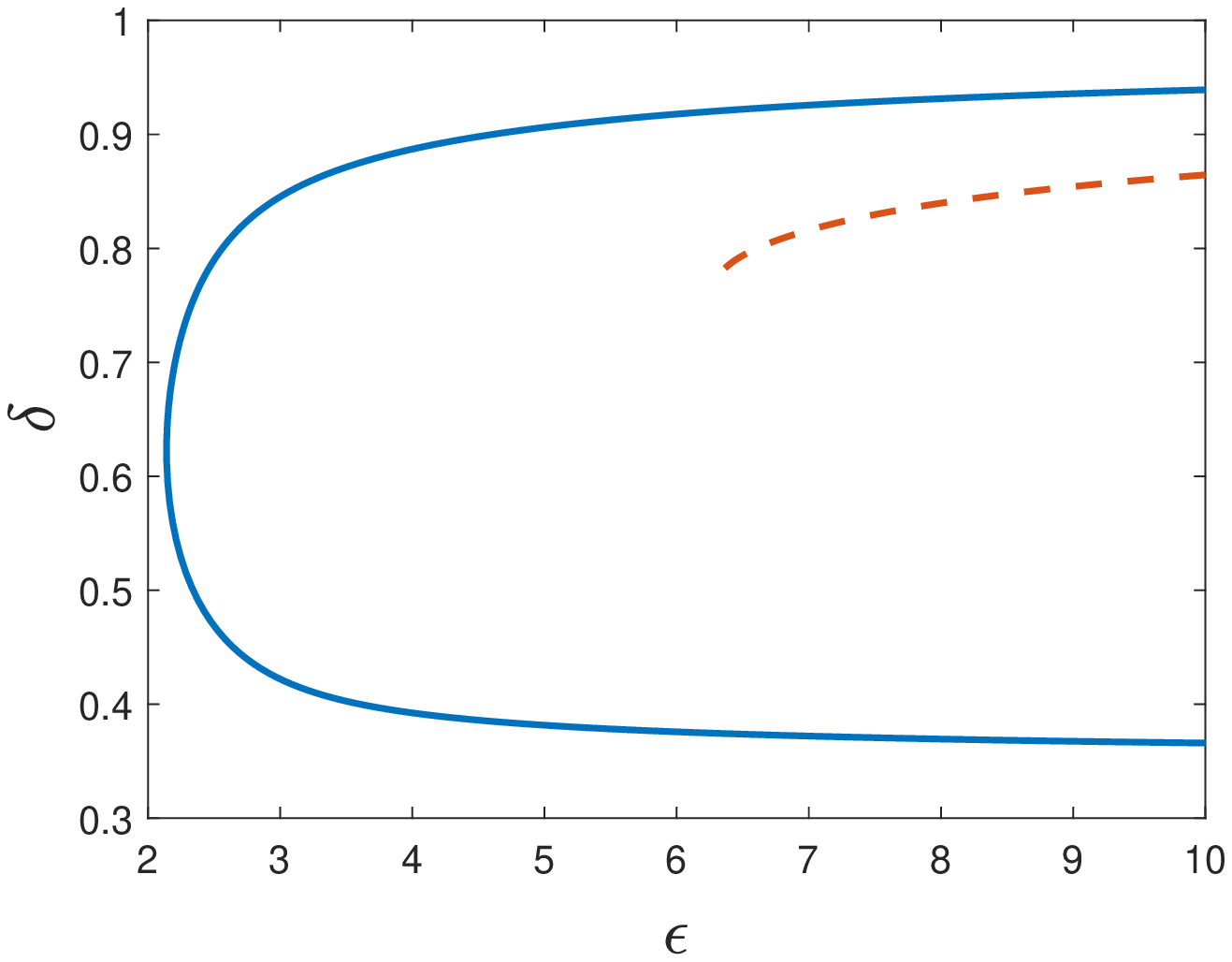}\label{fig1b}}\newline
\subfloat[][]{\includegraphics[width=7cm]{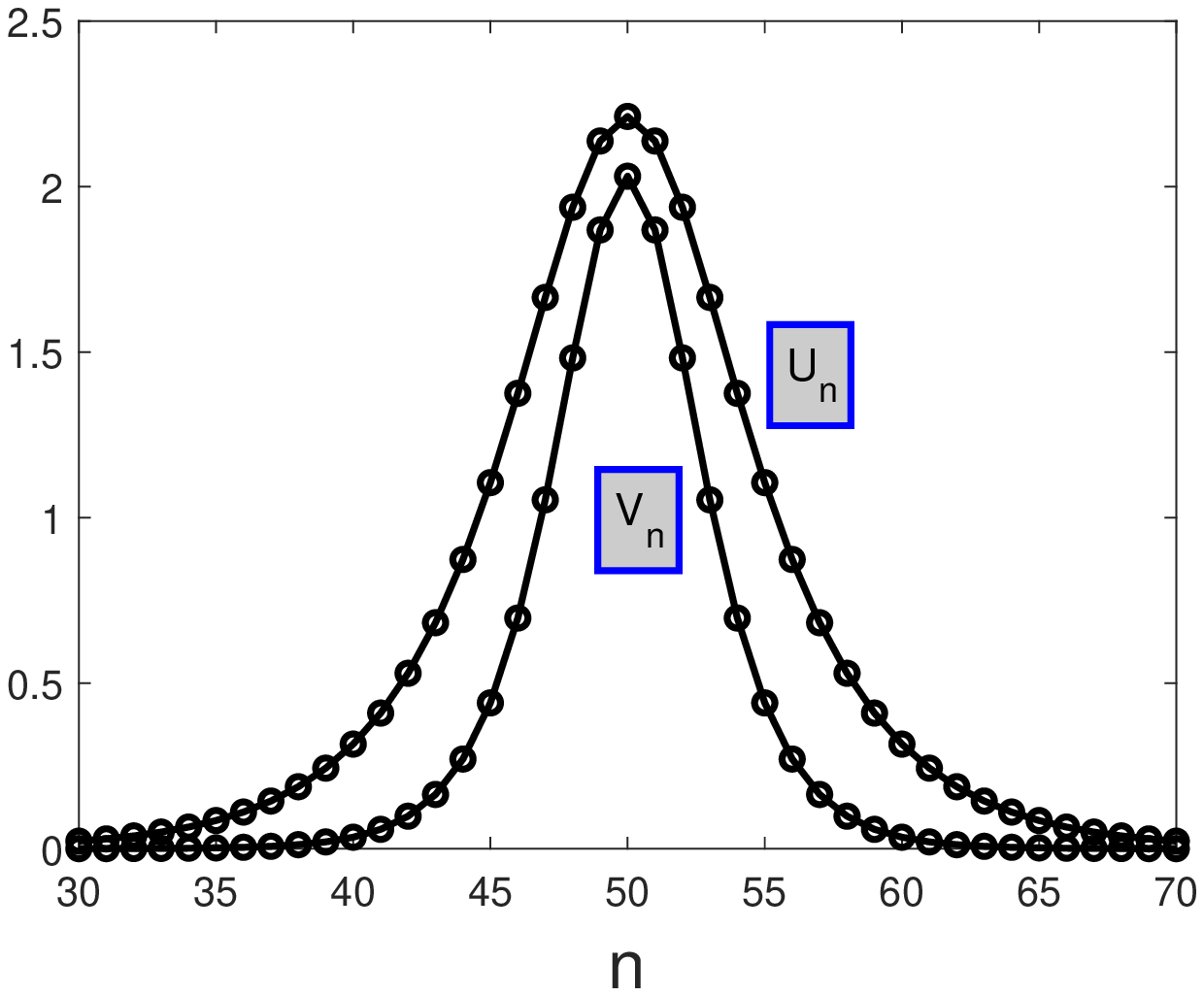}\label{fig1c}}
\caption{(a) A discrete embedded soliton (DES) for parameters of Eqs.\ (
\protect\ref{uv1}) and (\protect\ref{uv2}) $\protect\epsilon =10$, $q=1$, $%
\protect\omega =0.7$, $\protect\gamma _{1}=\protect\gamma _{2}=-0.05$, $%
\protect\delta \approx 0.9391$, and propagation constant $\protect\omega %
=0.7 $ [see Eq.\ (\protect\ref{stat})]. Circles and stars depict, severally,
a numerical solution of Eqs.\ (\protect\ref{U}), (\protect\ref{V}), and its
counterpart produced by the analytical approximation, based on Eqs.\ (%
\protect\ref{solc}), (\protect\ref{vaa}), (\protect\ref{vab}), (\protect\ref%
{solv}), and (\protect\ref{k}), (\protect\ref{Jc}). The location of DES
according to our approximation is at $\protect\delta \approx 0.8645$. (b) A
solid curve in the plane of $\left( \protect\epsilon ,\protect\delta \right)
$, along which the numerical solution produces DES states, with other
parameters fixed at the same values as in panel (a). The dashed curve shows
a similar result produced by the analytical approximation. (c) A numerically
found DES solution belonging to the lower branch in panel (b), for the the
same parameters as in (a), except for $\protect\delta \approx 0.3659$. In
panels (a) and (c), taller and lower profiles represent the FF and SH
fields, i.e., $U_{n}$ and $V_{n}$, respectively. DES families represented in
this figure are completely unstable. see Fig. \protect\ref{fig3} below.}
\label{sol}
\end{figure}

First, in Fig. \ref{sol}(a) we display an example of a numerically found
DES, obtained in the lattice built of $N=100$ sites, and its counterpart.
produced by means of the approximation based on Eqs.\ (\ref{solc}), (\ref%
{vaa}), (\ref{vab}), (\ref{solv}), and (\ref{k}), (\ref{Jc}), for values of
parameters in Eqs.\ (\ref{uv1}), (\ref{uv2}) $\epsilon =10$, $q=1$, $\gamma
_{1}=\gamma _{2}=-0.05$ (recall that $\gamma _{1,2}<0$ implies the
defocusing sign of the $\chi ^{(3)}$ nonlinearity) and propagation constant $%
\omega =0.7$. Because DES is a solution of the codimension-one type, for
these arbitrarily chosen parameters the numerical solution can be found only
for a specially selected value of the remaining one, \textit{viz}., the
relative strength of the lattice coupling for the SH wave, $\delta \approx
0.9391$, while the analytical approximation yeilds $\delta \approx 0.8645$
[recall that $\delta >0$ implies opposite signs of the SH and FF lattice
couplings in Eqs.\ (\ref{uv1}) and (\ref{uv2})]. 
At these values of the parameters, the gap in
the linear spectrum of the SH component, as given by Eq.\ (\ref%
{SHband-posdelta}), is $0.5<\omega <19.3$, hence $\omega =0.7$ indeed
belongs to the band (sitting close to its edge), confirming that the
discrete soliton is of the embedded type.

The use of the above approximation, corresponding to the continuum limit, is
relevant here, as $\epsilon =10$ is large enough for this purpose. Note also
that the relatively small value of the propagation constant, $\omega =0.7$,
makes the DES broad enough [as per Eq.\ (\ref{solc})], which additionally
justifies the use of the quasi-continuum formalism in this case. It is
remarkable that the analytical approximation produces a virtually exact DES
solution in Fig. \ref{sol}(a).

Fixing parameters $q=1$ and $\gamma _{1,2}=-0.05$, and propagation constant $%
\omega =0.7$ as above, the variation of $\epsilon $ makes it possible to
produce two branches of numerical solutions of Eqs.\ (\ref{U}), (\ref{V}) in
the $\left( \epsilon ,\delta \right) $ plane, as shown in Fig. \ref{sol}(b)
by the solid curve with a turning point at $\epsilon =\epsilon _{\min
}\approx 2.14$, at which the two branches merge. Thus, Fig.\ \ref{sol}(b)
shows that the continuation of the DES families does not exist if the
coupling constant falls below the minimum value. Note that, for the
currently fixed values, $q=1$ and $\omega =0.7$, the SH band, given by Eq.\ (%
\ref{SHband-posdelta}), amounts to $\epsilon \delta >0.1$. Obviously, the
entire solid curve in Fig.\ \ref{sol}(b) belongs to this area. It is also
worthy to note that both branches exist at $\delta >\delta _{\min }\approx
0.36$ [in particular, DESs do not exist at $\delta <0$, when the coupling
constants in the FF and SH sublattices have the same signs, see Eqs.\ (\ref%
{uv1}) and (\ref{uv2})].

The analytical approximation based on Eqs.\ (\ref{solc}), (\ref{vaa}), (\ref%
{vab}), (\ref{solv}), and (\ref{Jc}) produces a single existence branch,
shown by the dashed line in Fig. \ref{sol}(b), which cannot be extended
beyond the left end point, i.e., the approximation predicts, in a
qualitatively correct form, only one branch of the DES solutions. We assume
that this happens because ansatz \eqref{solc} is not sifficiently versatile
to capture the other branch of the solutions. A better ansatz may be tried,
but it would lead to an extremely cumbersome analysis.

\subsection{Narrow discrete embedded solitons}

\begin{figure}[tbph]
\centering
\subfloat[][]{\includegraphics[width=7cm]{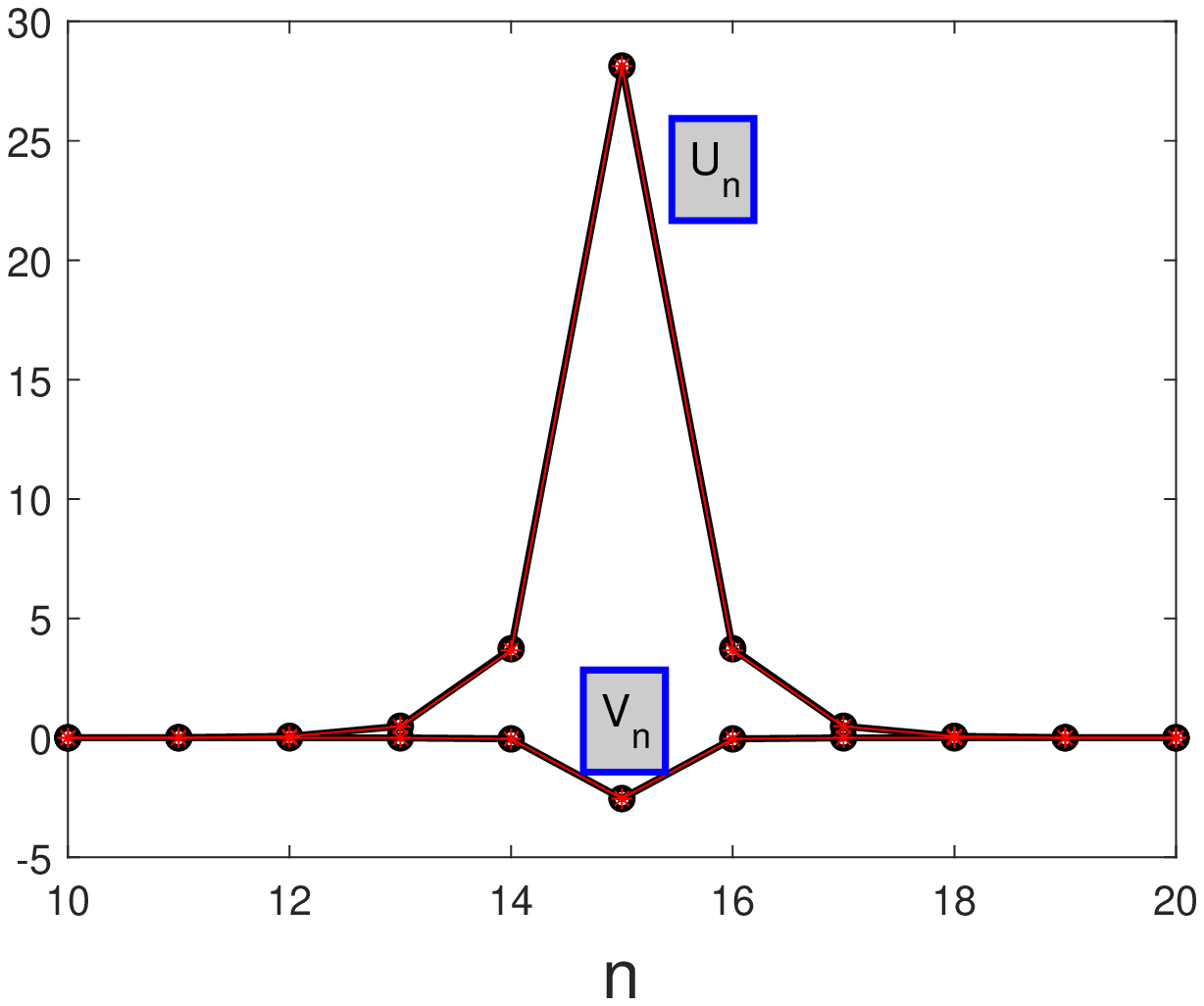}\label{fig2a}} %
\subfloat[][]{\includegraphics[width=7cm]{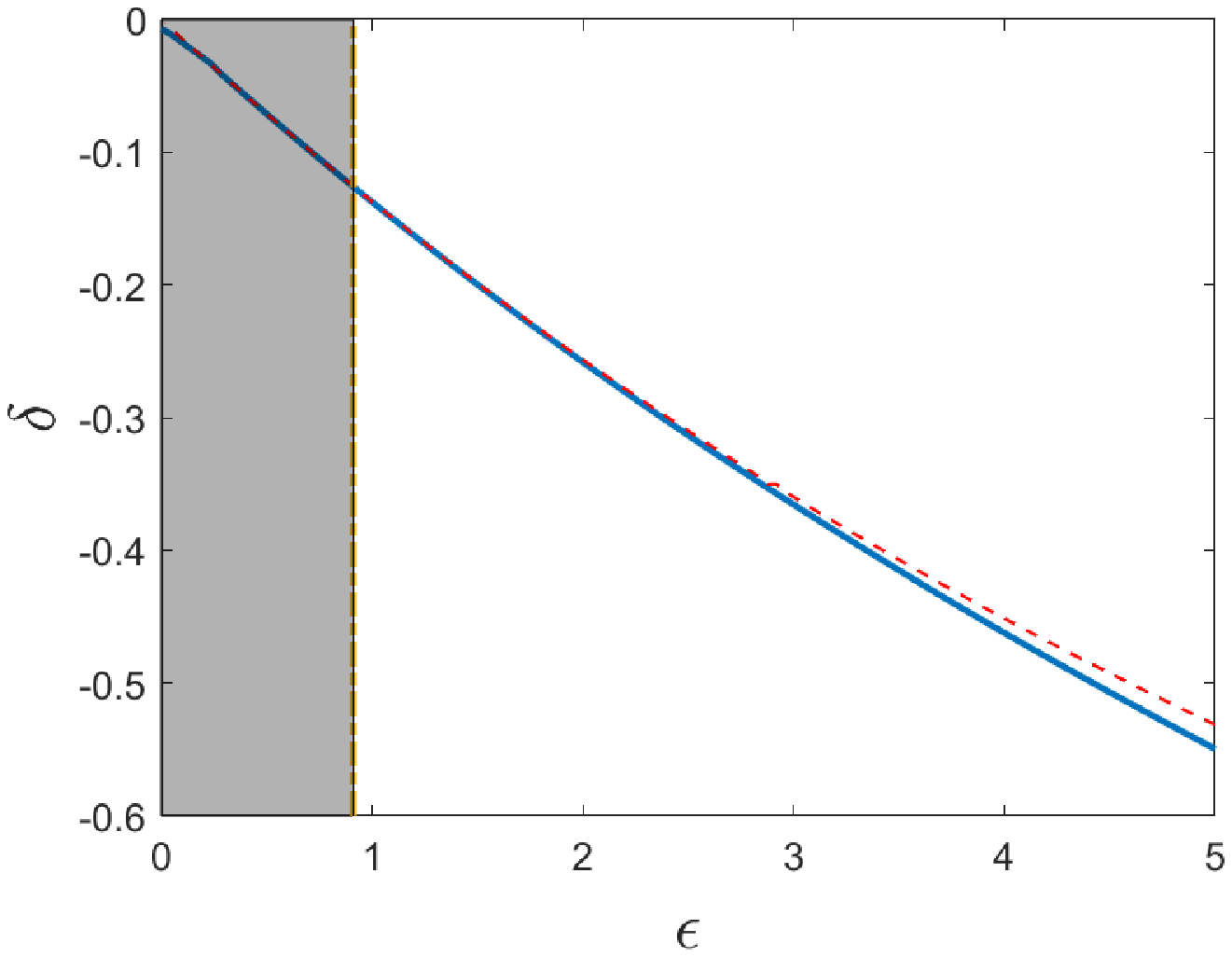}\label{fig2b}}
\caption{The same as Fig.\ \protect\ref{sol}(a,b), but for the weakly
coupled lattice with parameters $q=60$, $\protect\gamma _{1}=\protect\gamma %
_{2}=0.05$, and propagation constant $\protect\omega =29$. In panel (a), $%
\protect\epsilon =5$ and $\protect\delta \approx -0.5496$. The FF and SH
profiles, $U_{n}$ and $V_{n}$, are positive and negative ones. This DES is a
completely stable one, see Fig. \protect\ref{fig5} below. The shaded region
in panel (b) shows the domain of existence of regular (non-embedded)
solitons.}
\label{sol2}
\end{figure}

The above results were obtained for a situation close to the continuum
limit. Figure \ref{sol2} presents typical findings for the opposite case,
close to the anti-continuum limit, in the lattice composed of $N=30$ sites.
First, an example of a strongly discrete embedded soliton is displayed in
panel (a) for parameters of Eqs.\ (\ref{uv1}) and (\ref{uv2}) chosen as $%
\epsilon =5$, $q=60$, $\gamma _{1}=\gamma _{2}=0.05$, and propagation
constant $\omega =29$, while the value of coefficient $\delta $, at which
the nongeneric DES is found numerically, is $\delta \approx -0.5496$. Note
that the negative sign of $\delta $ implies identical signs of the
inter-site coupling in the FF and SH sublattices, on the contrary to the
case displayed above in Fig. \ref{sol}. At these values of the parameters,
the SH band, defined by Eq.\ (\ref{SHband-negdelta}), is $24.5<\omega <30$,
to which $\omega =29$ indeed belongs, sitting close to its edge.

The analytical approximation for the strongly discrete DES, based on Eqs.\ (%
\ref{ansatz}), (\ref{alpha}), and (\ref{select}) is also displayed, for the
same parameters, in Fig. \ref{sol2}(a). It is seen to be virtually identical
to the numerical counterpart.

In a still simpler approximation, which may be applied to a strongly
discrete soliton, its FF and SH amplitudes, $U_{0}$ and $V_{0}$, can be
found by completely neglecting the coupling of the central site, $n=0$, to
the adjacent ones, $n=\pm 1$ \cite{Aubry}. In this limit, Eqs.\ (\ref{U}), (%
\ref{V}) yield, in the expectation of having $U_{0}^{2}\gg V_{0}^{2}$, the
following values:%
\begin{equation}
V_{0}\approx -1/\left( 8\gamma _{2}\right) =-2.5,U_{0}\approx \sqrt{\left(
\omega -V_{0}\right) /\gamma _{1}}\approx 25.1,
\end{equation}%
which are close enough to ones observed in Fig. \ref{sol2}(a).

A family of DES modes, produced by the numerical solution of Eqs.\ (\ref{U}%
), (\ref{V}) for the same parameters as in Fig. \ref{sol2}(a), i.e., $q=60$,
$\gamma _{1,2}=0.05$, and $\omega =29$, but with a varying coupling constant
$\epsilon $, and $\delta $ adjusted accordingly, is displayed in Fig. \ref%
{sol2}(b), along with its counterpart produced by the above-mentioned
analytical approximation, based on Eqs.\ (\ref{ansatz}), (\ref{alpha}), and (%
\ref{select}), It is clearly seen that the approximation is very accurate in
this case, even for relatively large values of $\epsilon $, for which the
applicability of the anti-continuum formalism is not evident. Unlike the
family shown above in Fig. \ref{sol}, with identical signs of the FF and SH
components, the present one always features opposite signs.

Lastly, we note that, for the parameters chosen in Fig. \ref{sol2}(b), Eq.\ (%
\ref{SHband-negdelta}) defines the SH band as the area with $|\delta
|\epsilon >0.5$. This condition means that the family presented in this
figure belongs to the band at $\epsilon >\epsilon _{0}\approx 0.9098$, while
its extension to $\epsilon <\epsilon _{0}$ immerses into the bandgap, as a
family of regular solitons (non-embedded ones). In the figure, the existence
region of the regular solitons is shaded.



\subsection{Stability and evolution of discrete embedded solitons}

\begin{figure}[tbph]
\centering
\subfloat[][]{\includegraphics[width=7cm]{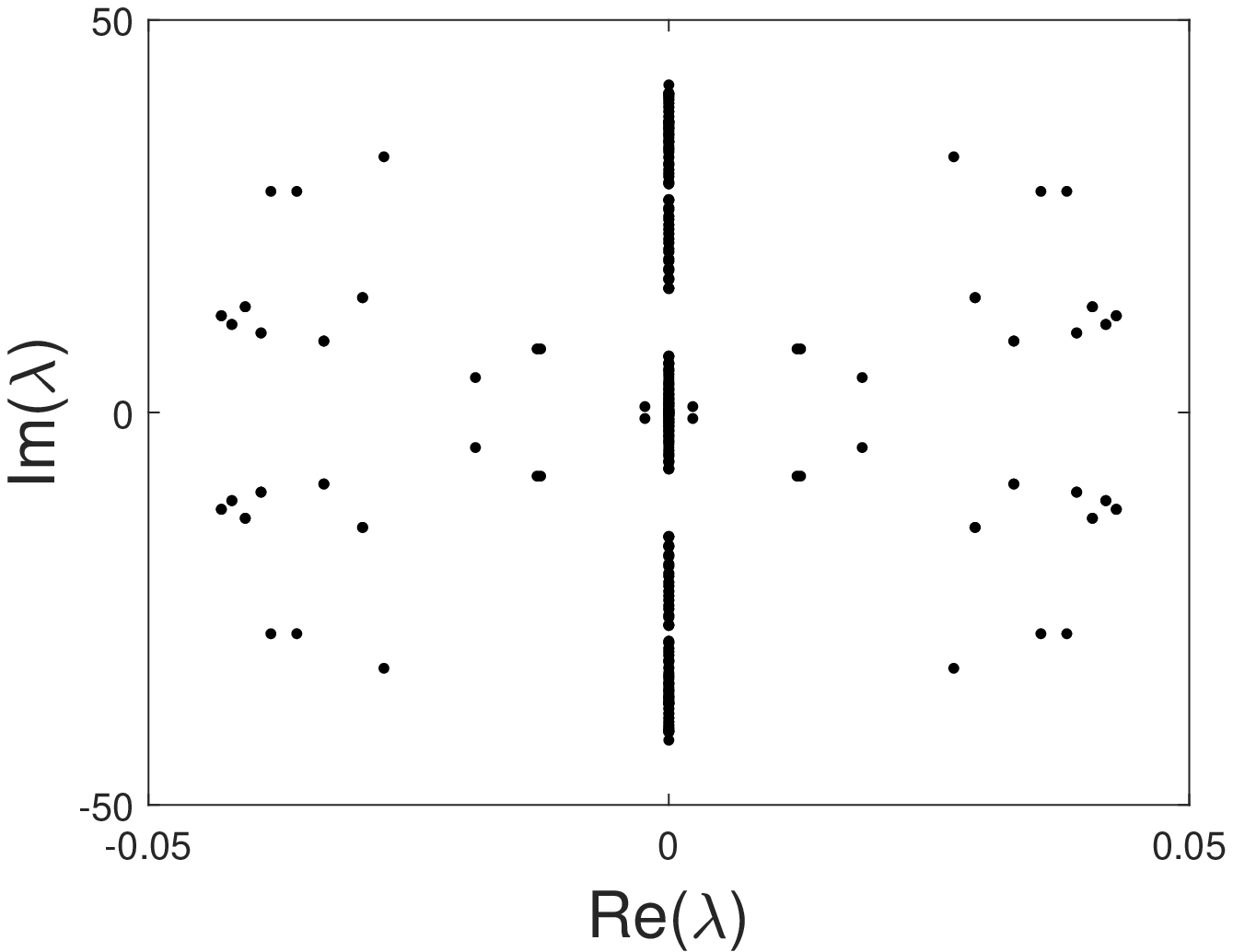}\label{fig3a}} %
\subfloat[][]{\includegraphics[width=7cm]{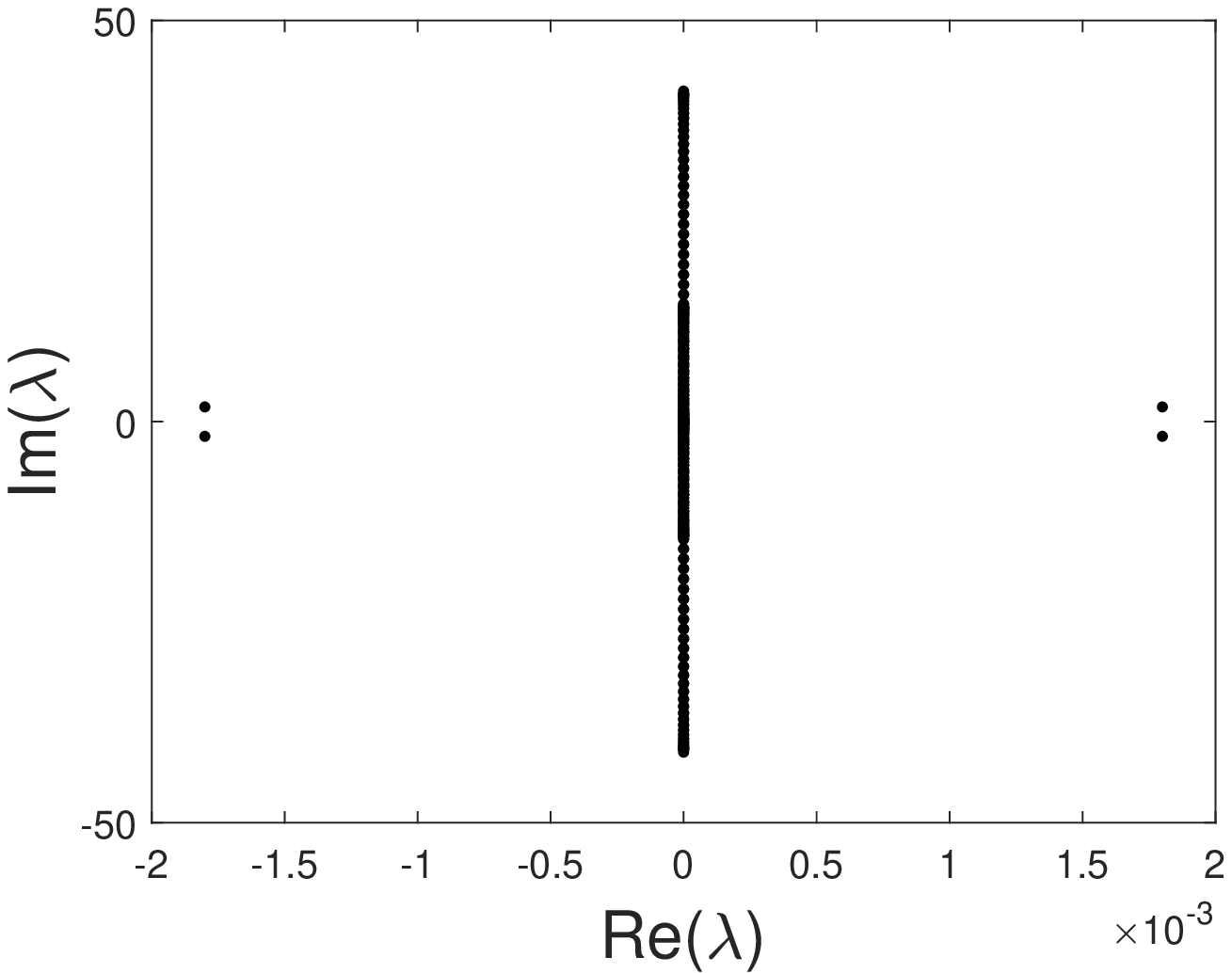}\label{fig3b}}
\caption{Panels (a) and (b) show spectra of complex stability eigenvalues
for DES displayed in Figs.\ \protect\ref{fig1a} and \protect\ref{fig1c},
respectively.}
\label{fig3}
\end{figure}

We address the stability of DESs by numerically solving the eigenvalue
problem \eqref{evp}. First, we consider the case of opposite signs of the
coupling constant in the SH and FF lattices, i.e., $\delta >0$ in Eq. (\ref%
{uv2}). The first result is that the solution branches in Fig.\ \ref{fig1b}
are completely unstable. As an illustration, the spectrum of the eigenvalues
displayed in Figs.\ \ref{fig1a} and \ref{fig1c} is shown in Fig.\ \ref{fig3}%
. In these cases, the oscillatory instability is accounted for by quartets
of complex eigenvalues.

\begin{figure}[tbph]
\centering
\subfloat[][]{\includegraphics[width=7cm]{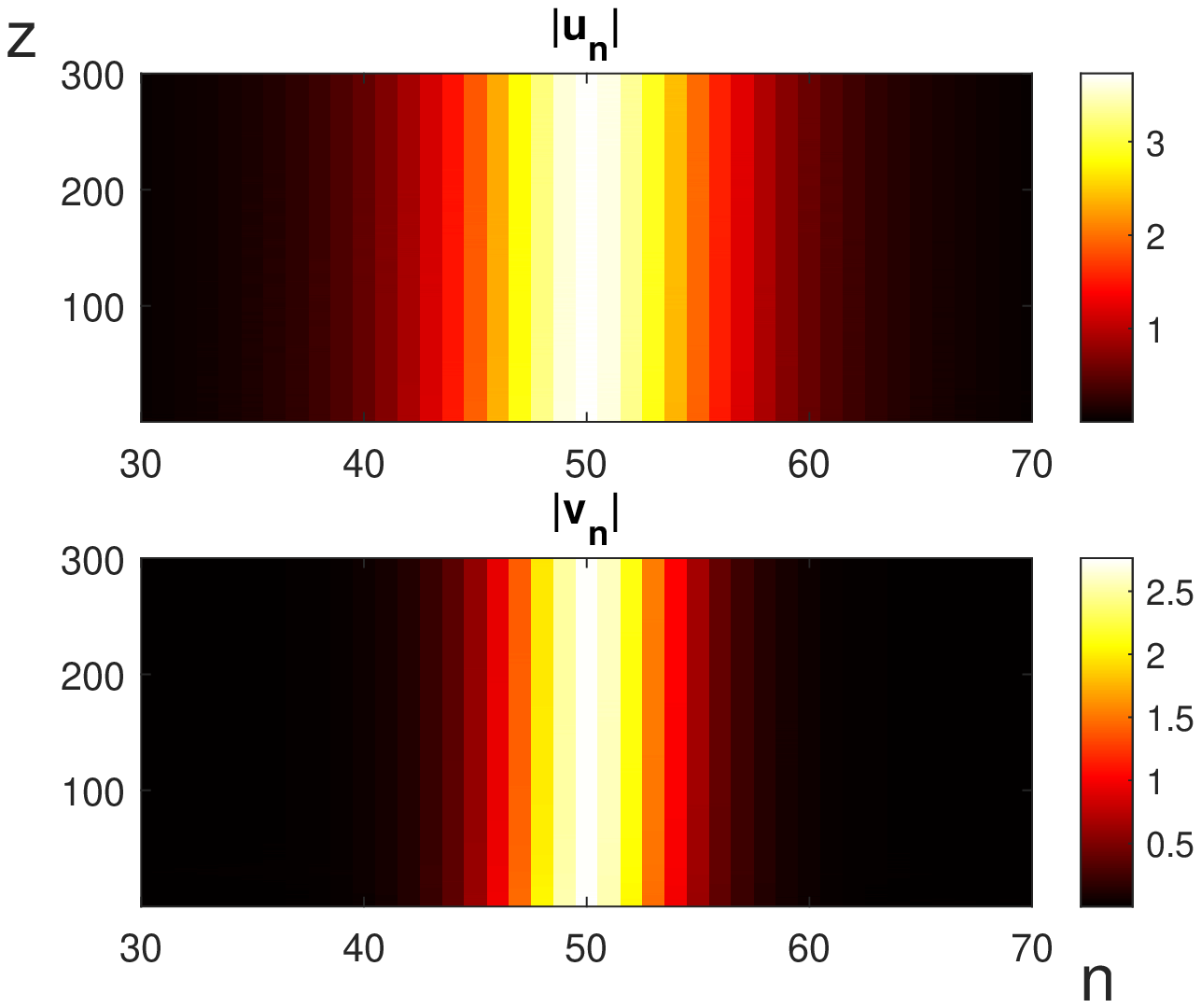}\label{fig4a}} %
\subfloat[][]{\includegraphics[width=7cm]{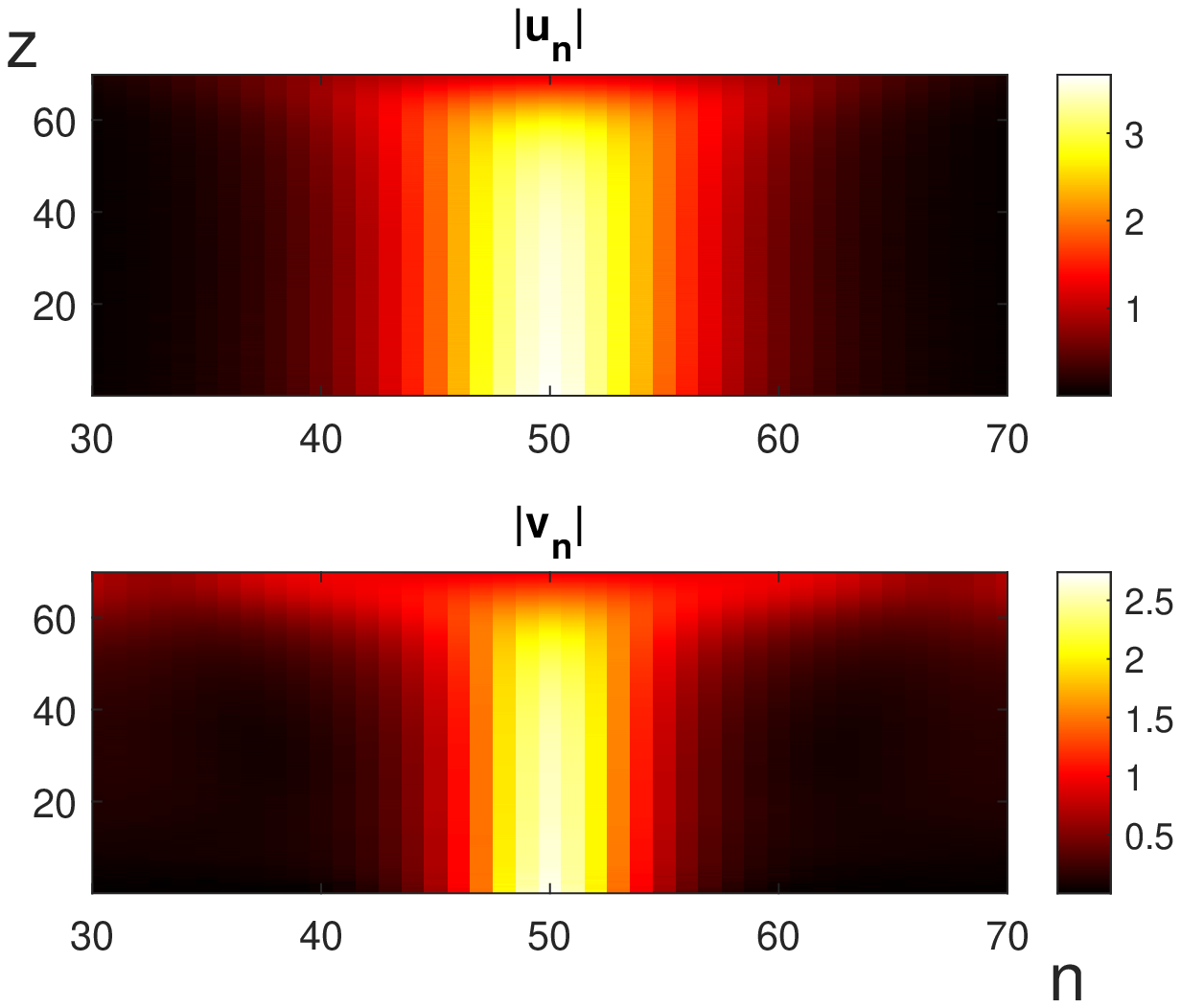}\label{fig4b}}
\caption{The evolution of DES from Fig.\ \protect\ref{fig1a}(a) under the
action of complex random-noise perturbations with maximum modulus amplitude $2\times10^-2$; and (b) after being initially
perturbed by rescaling its shape, according to $u_{n}(t=0)=0.99\times U_{n}$%
, $v_{n}(t=0)=0.99\times V_{n}$.}
\label{fig4}
\end{figure}

Having established the instability of the DESs shown in Fig. \ref{fig1a}, we
display simulations of their dynamics, under the action of two different
perturbations, in Fig.\ \ref{fig4}. First, Fig.\ \ref{fig4a} demonstrates
that the soliton, in spite of the presence of many unstable eigenvalues in
its perturbation spectrum, as seen in Fig.\ \ref{fig3a}, is actually quite
robust in direct simulations which include small random perturbations in the
initial conditions. In this connection, we note that, with $\epsilon =10$
and $\delta $ close to $1$ (the values used in Fig. \ref{sol}), and the
soliton's width $W\sim 10$ (as seen in the same figure), the characteristic
diffraction (Rayleigh) length is $Z_{\mathrm{diffr}}\sim W^{2}/\epsilon \sim
10$, i.e., the total propagation distance displayed in Fig. \ref{fig4}(a) is
$\sim 30$ diffraction lengths, which is actually a very large value in a
real experiment \cite{Torruellas}-\cite{review3}. On the other hand, if the
initial state is perturbed merely by multiplying it by factor $0.99$, the
DES originally stays nearly intact, over the distance $Z\simeq 60\sim 6Z_{%
\mathrm{diffr}}$, and then quickly decays, as shown in Fig.\ \ref{fig4b}.
The latter dynamical scenario is typical for the \textquotedblleft
one-sided" semistability of embedded solitons \cite{yang99}, initiated by
slow sub-exponential growth of perturbations, which eventually leads to fast
destructiom of the solitons. The perturbed evolution of the DES displayed in
Fig.\ \ref{fig1c} exhibits similar dynamical behavior (not shown here in
detail).

\begin{figure}[tbph]
\centering
\subfloat[][]{\includegraphics[width=7cm]{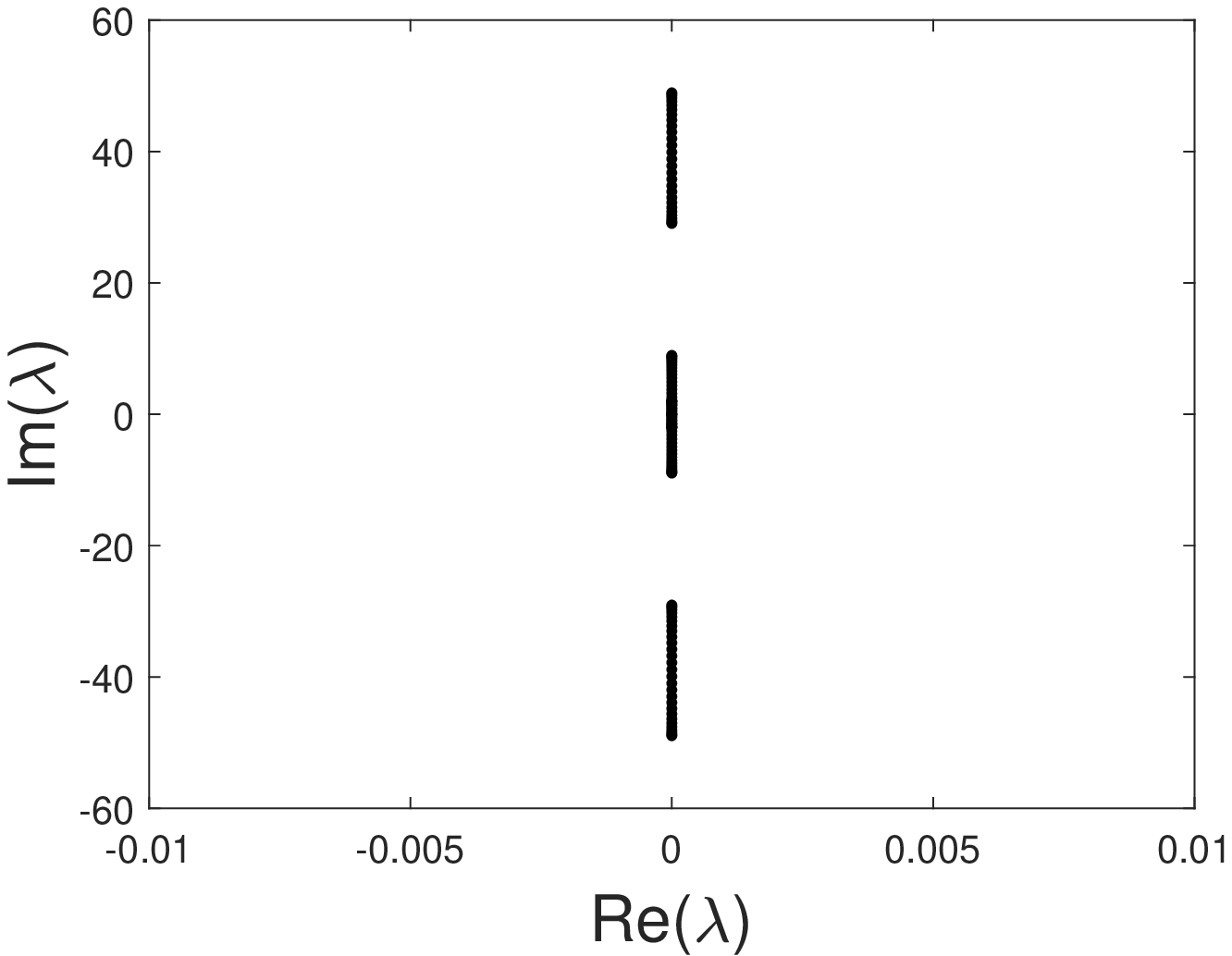}\label{fig5a}} %
\subfloat[][]{\includegraphics[width=7cm]{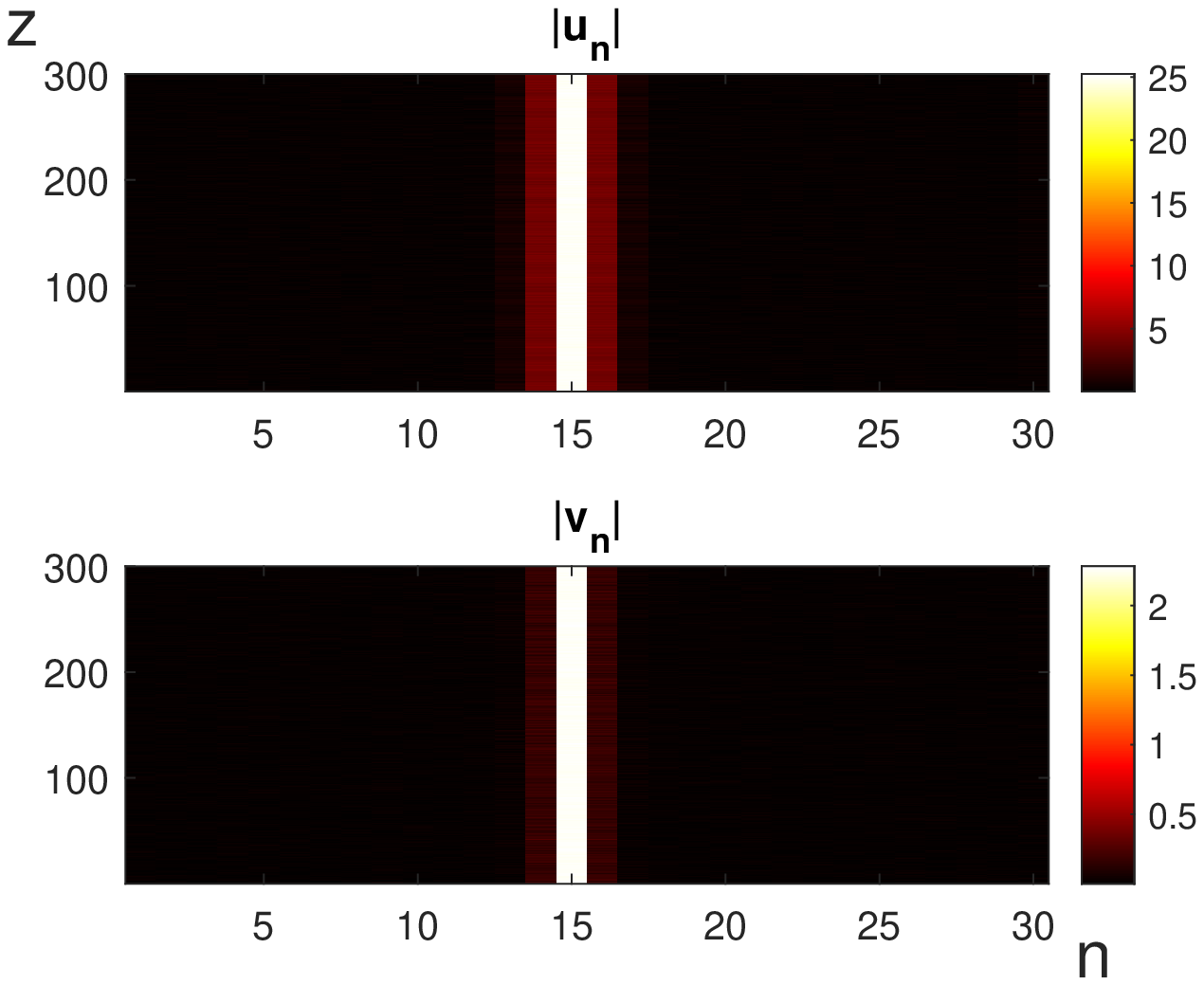}\label{fig5b}}
\caption{(a) The spectrum of perturbation eigenvalues for the DES from Fig.\
\protect\ref{fig2a}. (b) The simulated evolution of the same soliton, with
an initial perturbation in the form of $u_{n}(t=0)=0.9\times U_{n}$, $%
v_{n}(t=0)=0.9\times V_{n}$. This, relatively strong, perturbation leaves
the soliton completely stable, unlike a much weaker initial scaling
perturbation which destroys the DES in Fig. \protect\ref{fig4}(b).}
\label{fig5}
\end{figure}

Next, we consider the case of $\delta <0$, i.e., opposite signs of the
intersite linear couplings in the FF and SH lattices in Eqs. (\ref{uv1}) and (\ref{uv2}). For the DES shown in Fig.\ \ref{fig2a}, its linear-perturbation spectrum is plotted in Fig.\ \ref{fig5a}, which does not include unstable eigenvalues (the spectrum is purely imaginary). This is the case with all DESs along the existence curve plotted in Fig.\ \ref{fig2b}. Direct simulations of the
evolution of the soliton subjected to the scaling perturbation, similar to
that which led to the destruction of the semistable soliton in Fig. \ref%
{fig4}(b), but actually much stronger (with the initial scaling factor $0.9$
instead of $0.99$), demonstrate perfect stability. Under the action of
random perturbations, the same soliton is perfectly stable as well (not
shown here in detail). Thus, the system of equations (\ref{uv1}) and (\ref%
{uv2}) with $\delta <0$ produces fully stable DESs, which were not reported
before..

\section{Conclusion}

\label{conc}

In this work, we have revisited the problem of the creation of DESs
(discrete embedded solitons) in the model of the light propagation in an
array of tunnel-coupled waveguides with the combined on-site quadratic and
cubic nonlinearities. Two analytical approximations have been developed,
based on the averaging method, which apply, severally, to broad and narrow
DESs. Systematic results are obtained by means of numerical calculations.
Both analytical and numerical methods produce DESs of different types -- in
particular, with identical or opposite signs of the FF
(fundamental-frequency) and SH (second-harmonic) components. The analytical
approximation yields very accurate results, in comparison to their
numerically found counterparts, for the most interesting case of strongly
discrete embedded solitons. In that case, the DES family crosses the border
of the spectral propagation band and extends, as a branch of regular
solitons, into the semi-infinite gap. The analytical approximations
developed here for the DESs may be applied to other physically relevant
discrete models. Stability of the DESs has been investigated numerically.
While it is generally known that DESs are semi-stable objects, which we
observed as well here, in the case of FF and SH components having intersite
coupling coefficients with opposite signs, we demonstrate the existence of
remarkably stable DESs, which, to the best of our knowledge, has not been
reported before.

A challenging direction for the development of the present analysis is to
perform it for two-dimensional lattices with the $\chi ^{(2)}:\chi ^{(3)}$
nonlinearity, with the aim to seek for two-dimensional DESs.

\section*{Competing interests}

The authors declare that they have no known competing financial interests or personal relationships that could have appeared to influence the work reported in this paper.

\section*{CRediT author statement}

\textbf{Hadi Susanto}: Conceptualization, Methodology, Formal analysis, Investigation, Writing - Original Draft, Visualization. \textbf{Boris Malomed}: Conceptualization, Methodology, Writing - Review \& Editing.

\section*{Acknowledgements}

HS thanks Mahdhivan Syafwan for fruitful discussions at the early stage of
this work.
The work of BAM is supported, in part, by Israel Science Foundation through
grant No. 1286/17. This author appreciates hospitality of Department of
Mathematical Sciences at the University of Essex (UK).

\end{document}